\documentclass[Times,4pt,aps,pra,twocolumn,showpacs,amsmath,amssymb,floatfix,footinbib,superscriptaddress]{revtex4-1}
\usepackage{amsmath,natbib,graphicx,subfigure,amssymb,graphics,amsmath,mathrsfs,CJK,color}
\usepackage{multirow,fancyhdr,color,bm,tabularx,psfrag,dcolumn}
\usepackage[dvipdfm,colorlinks=true,linkcolor=blue,urlcolor=blue,citecolor=blue]{hyperref}
\usepackage[mathlines]{lineno}
\def\footnoterule{\kern -1mm \hrule width 6cm \kern 2.2mm}
\begin{document}

\title{ Photovoltaic properties evaluated by its thermodynamic evolution in a double quantum dot photocell}
\author{Sheng-Nan Zhu }
\affiliation{Center for Quantum Materials and Computational Condensed Matter Physics, Faculty of Science, Kunming University of Science and Technology, Kunming, 650500, PR China}
\affiliation{Department of Physics, Faculty of Science, Kunming University of Science and Technology, Kunming, 650500, PR China}

\author{Shun-Cai Zhao}
\email[Corresponding author: ]{zhaosc@kmust.edu.cn}
\affiliation{Center for Quantum Materials and Computational Condensed Matter Physics, Faculty of Science, Kunming University of Science and Technology, Kunming, 650500, PR China}
\affiliation{Department of Physics, Faculty of Science, Kunming University of Science and Technology, Kunming, 650500, PR China}

\author{Lin-Jie Chen }
\affiliation{Center for Quantum Materials and Computational Condensed Matter Physics, Faculty of Science, Kunming University of Science and Technology, Kunming, 650500, PR China}
\affiliation{Department of Physics, Faculty of Science, Kunming University of Science and Technology, Kunming, 650500, PR China}
\begin{abstract}
Obtaining the physical mechanism of  photoelectric transfer in quantum-dot (QD) photocells may be one strategy to boost the photovoltaic conversion efficiency. In this work, we attempted to formulate a novel theoretical approach to evaluate photocells¡¯ performance via evaluating their thermodynamic evolution during the photoelectric conversion process in a double quantum dot (DQD) photocell model. Results demonstrate that the thermodynamic-related quantities can reflect the law of photovoltaic dynamics, i.e., the photoelectric transfer properties can be evaluated by the heat currents indirectly. The merit of this work not only expands our understanding of the physical law of heat currents in the photoelectric transport process, but it may also propose a new method for optimizing photoelectric conversion efficiency in a DQD photocell.
\begin{description}
\item[Keywords]{Thermodynamic evolution; photovoltaic properties; photoelectric transfer}
\end{description}
\end{abstract}


\maketitle
\section{Introduction}
In nonequilibrium statistical physics, the energy transport law is a fundamental issue that has been studied over and over again\cite{li_2012_icolloquiumi,li_2021_influence,1-xu2022Photosynthetic,4-zhongzhao2021}.
In particular, the study of energy transport at quantum scale has attracted the attentions from theoretical and experimental researchers\cite{gardiner_1991_quantum,heinzpeterbreuer_2006_the}. For example, the energy transport in the photocells converting photon energy from the solar into electric energy, interests quantum physics\cite{scully2011quantum} due to the efficient conversion for mitigating the gradual depletion of fossil fuels. And many theoretical works have revealed that Fano-induced coherence\cite{svidzinsky_2011_enhancing,chen_2020_radiative}, noise-induced quantum coherence\cite{scully2011quantum}, delocalized quantum states of interacting dipoles\cite{zhang_2015_delocalized} can reduce the radiative recombination loss of a photocell and then enhancing the efficiency of photocells\cite{zhong_2021_photovoltaic}.

Thermodynamics deals with the evolution of a system, often in relation to its ambient environment, describing the general dynamical laws under universal laws independent of microscopic details\cite{1979The,cuetara_2016_quantum,PhysRevLett.122.150603}. Among its four laws, the first and second laws of thermodynamics plays an important role in our understanding of complex physical systems\cite{spohn_1978_irreversible}. The first law of classical thermodynamics states the energy conservation\cite{seifert_2012_stochastic}, the second law\cite{evans_1993_probability} dictates the irreversible evolution of all physical and chemical processes related to thermal motion in limited space and time. However, the relationship has not been fully and clearly revealed between energy transport and thermodynamic evolution, at least in the photovoltaic process. Therefore, we attempted to formulate a novel theoretical approach to evaluate photovoltaic performance, through discussing thermodynamic-related physical quantities during the photovoltaic process in this work via a proposed DQD photocell model.

With this aim in mind, this paper is organized as follows. In Section 2, we briefly review the quantum thermodynamics of an open quantum system. And then we examine the photovoltaic properties of the proposed DQD photocell model based on the master equation approach in Section 3. The results and analysis will be presented in Section 4, we calculate the quantum thermodynamic and photovoltaic characteristic quantities of the DQD photocell, and compare the thermodynamic and photovoltaic evolution process under the same parameter conditions. Finally, in Section 5 we summarize our results with some discussion.

\section{General Formalism}

In the thermodynamic laws, the first law of classical thermodynamics states the energy conservation, $\delta E$=$\delta W$ + $\delta Q$. The changes in internal energy  $\delta E$ can be distributed by the work $\delta W$ performed on the system and heat $\delta Q$ transport to the system\cite{1979The,PhysRevA.74.063823}. If an open quantum system weakly coupled to the bath $\alpha$, the time-dependent internal energy of the system is given by $E(t)$=$Tr[\hat{H}_{T}(t)\hat{\rho}(t)]$ with a total Hamiltonian $\hat{H}_{T}(t)$=$\hat{H}_{S}+\hat{H}_{1}(t)$ \cite{dong_2021_thermodynamic,Weimer_2008}, in which \(\hat{H}_{S}\) is the system Hamiltonian and \(\hat{H}_{1}(t)\)  is the interaction between the system and the ambient thermal environment. Its derivative with respect to time gives rise to the first law of quantum thermodynamics as follows,

\begin{equation}
\dot{E}(t)=Tr[\dot{\hat{H}}_{T}(t)\hat{\rho(t)}]+Tr[\hat{H}_{T}(t)\dot{\hat{\rho}}(t)]=\dot{W}(t)+ \dot{Q}(t)  \label{1}
\end{equation}

\noindent Here $Tr[\dot{\hat{H}}_{T}(t)\hat{\rho(t)}]$=$\dot{W}(t)$  is the power transported to the system by external forces, $Tr[\hat{H}_{T}(t)\dot{\hat{\rho}}(t)]$=$\dot{Q}(t)$  is defined as the heat current from the bath into the system\cite{quan_2007_quantum,1980Thermodynamic}. If the system Hamiltonian \(\hat{H}_{S}\) is independent of time, $\hat{H}_{T}(t)$=$\hat{H}_{1}(t)$. The second law of thermodynamics describe the thermodynamic evolution of the system. Two equivalent formulations state this law within equilibrium thermodynamics. One is the Clausius inequality, which states that the entropy production rate of the system is non-negative\cite{cuetara_2016_quantum,Spohn1978Entropy,kadison_1965_mathematical} during the transition between two equilibrium states,

\begin{equation}
\dot{\sigma}=\frac{dS}{dt}-\sum_{\alpha}\beta_{\alpha}\dot{Q}_{\alpha} \ge 0  \label{2}
\end{equation}

\noindent where $\beta_{\alpha}$=$\frac{1}{k_{B}T_{\alpha}}$, and $k_{B}$ is the Boltzmann constant. The von Neumann entropy S\cite{2019The,li2017new} in Eq.(\ref{2}) is given by $S$=$-Tr[\rho(ln\rho)]$, thus

\begin{equation}
\frac{dS}{dt}=-Tr[\dot{\rho}(ln\rho)+\rho(ln\dot{\rho})]=-Tr[\dot{\rho}(ln\rho)] \label{3}
\end{equation}

\noindent The above equation takes advantage of population normalization conditions, $Tr[\dot{\rho}=0]$. The other formulation, referred to as the free energy inequality, declares that \(W\)-\(\Delta F\) \(\geq\) 0 during the transition between two equilibrium states, where \(W\) is the work performed on the system and \(F\)=\(E\)-\(T S\) is the free energy (here \(E\) states the internal energy, the von Neumann entropy S with the temperature \(T\)).
Since quantum dots are quasi-zero-dimensional artificial quantum structures with discrete electronic states, so energy transport in the quantum dots can be easily regulated\cite{2017General,doi:10.1021/nl101490z, doi:10.1021/ar3001958}. These characteristics make QDs be used as a perfect energy filter and efficient light harvesting device\cite{2019PbS,Fan029,Zhao_2019}. Moreover, lots of work carried out on the thermodynamic properties of QD photocells \cite{li_2021_influence,scully2011quantum, dong_2021_thermodynamic}.

However, few studies have focused on the thermodynamic evolution of photovoltaic processes in QD photocells. This issue will help to reveal their quantum photoelectric transport properties  from the perspective of thermodynamics. Therefore, we attempted to formulate a novel theoretical approach to evaluate DQD photocells performance in this work. The heat currents between the donor and acceptor in the DQD photocells were shown to be responsible for this performance. Here, a donor-acceptor model will be proposed in the DQD photocell, which is driven by photon pulses and the thermodynamic properties are shown during photovoltaic evolution process.

\section{Physical Model and Solutions}

\begin{figure}[htp]
{\includegraphics[scale=.45]{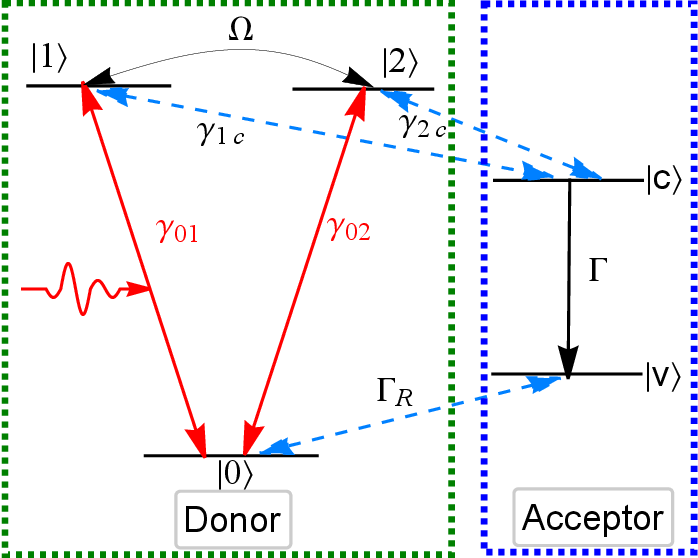}}\hspace{0in}%
\caption{(Color online) Schematic diagram of photon pulse-driven donor-acceptor DQD photocell model. The driving photon pulse is represented by the red wavy arrow. We denote the valence band (VB) states by level $|0\rangle$, and the conduction band (CB) states by levels $|1\rangle$, $|2\rangle$ in the DQD photocell. The photon pulse drives the electric transition between CBs ($|1\rangle$,$|2\rangle$) and VB ($|0\rangle$) at a rates of $\gamma_{0i}(i=1,2)$. States $|c\rangle$ and $|v\rangle$ are connected to an external terminal at the rate $\Gamma$ standing for the external load or electrical resistance. Transitions $|i\rangle(i=1,2)$ $\leftrightarrow$ $|c\rangle$, $|v\rangle$ $\leftrightarrow$ $|0\rangle$ driven by the ambient thermal phonons, are accompanied by the election transport between the donor and acceptor with relaxation rates $\gamma_{ic}(i=1,2)$ and $\Gamma_{R}$.}
\label{Fig.1}
\end{figure}

As shown in Fig.\ref{Fig.1}, the DQD photocell we consider is a 5-level quantum system composed of a donor and an acceptor. The cyclic operation of the donor-acceptor DQD photocell can be performed with the sequence as follows:
(1) The donor absorbs incoming photons and the electrons become excited with the transitions from the valence band (VB) state $|0\rangle$, to the conduction band (CB) states $|1\rangle$ and $|2\rangle$. (2) The phonon vibration makes the excited electrons at the donor transfer to the acceptor state $|c\rangle$. (3) The acceptor is coupled to an external load and the electric current represented by the transition decays from the state $|c\rangle$ to the state $|v\rangle$. (4) The electrons in the state $|v\rangle$ of the acceptor return to the VB state $|0\rangle$ of the donor by a non-radiative decay. Consider a strong Coulomb interaction, with an extra electron in the left or right dot. The total DQDs photocell system can be modeled by the Hamiltonian,

\begin{equation}
\hat{H}_{T}=\hat{H}_{S}+\hat{H}_{photon}+\hat{H}_{V}.\label{4}
\end{equation}

In this donor-acceptor DQD model (Illustrated in Fig.\ref{Fig.1}), the systematic Hamiltonian, $\hat{H}_{S}$=$\hat{H}_{A}+\hat{H}_{D}$ contains two parts: the acceptor Hamiltonian $\hat{H}_{A}$ and the donor Hamiltonian $\hat{H}_{D}$. (For the sake of convenience, we set $\hbar$=$1$). And $\hat{H}_{A}$ is read as,

\begin{equation}
\hat{H}_{A}=\frac{\omega_{0}}{2}(|c\rangle\langle c|-|v\rangle\langle v|)  \label{5}
\end{equation}

\noindent where $\omega_{0}$ is the transition frequency between the acceptor states $|c\rangle$ and $|v\rangle$. The donor in this DQD model is driven by photon pulses, hence, $\hat{H}_{D}$ with the expressions are listed as follows,

\begin{align}
\hat{H}_{D}=\hat{H}^{D}_{0}+\hat{H}^{D}_{1}(t)\,\label{6}
\end{align}

\begin{eqnarray}
&\hat{H}^{D}_{0}=&\frac{\varepsilon}{2}(|1\rangle\langle 1|-|2\rangle\langle 2|)+\Omega(|1\rangle\langle 2|+|2\rangle\langle 1|)\, \label{7}\\
&\hat{H}^{D}_{1}(t)=&i(g^{\ast}_{1}(t)\sqrt{\gamma_{01}}\sigma^{-}_{1}-g_{1}(t)\sqrt{\gamma_{01}}\sigma^{\dag}_{1})\nonumber\\
                  &&+i(g^{\ast}_{2}(t)\sqrt{\gamma_{02}}\sigma^{-}_{2}-g_{2}(t)\sqrt{\gamma_{02}}\sigma^{\dag}_{2}) \label{8}
\end{eqnarray}

\noindent Among Eq.(\ref{6}), $\hat{H}^{D}_{0}$ describes the eigenenergy Hamiltonian of donor DQD photocell, and $\hat{H}^{D}_{1}(t)$ describes the interaction between donor and incident photon pulses. $\varepsilon$ denotes the eigenstate energy, and $\Omega$ is the tunneling rate which describes the tunneling effect between two quantum dots in Eq.(\ref{7}). It is well known that the tunneling rate is a characteristic that reflects the fundamental physics of DQD. However, extensive research\cite{lira_2022_enhanced,4-zhongzhao2021} has demonstrated the physical relationship between tunneling rate and photovoltaic performance in the DQD photocell. Therefore, we will focus on how to evaluate the photovoltaic properties through the thermodynamic evolution, i.e., some thermodynamic parameters in this study, and the tunneling rate will be assigned a constant for the discussion. $\sigma^{-}_{i}$ ($\sigma^{\dag}_{i}$) is a lowering operator (raising operator) that describes the corresponding transition process in Eq.(\ref{8}), $\sigma^{-}_{1}$=$|0\rangle\langle 1|$, $\sigma^{\dag}_{1}$=$|1\rangle\langle 0|$, $\sigma^{-}_{2}$=$|0\rangle\langle 2|$, $\sigma^{\dag}_{2}$=$|2\rangle\langle 0|$. Here, we consider a Gaussian pulse shape \cite{loudon2010the,wang2011efficient} photon pulse $g_{i}(t)$ and the transition rate $\gamma_{0i}$ coupling to the donorsin Eq.(\ref{8}). $\hat{H}_{photon}$ in Eq.(\ref{4}) represents the Hamiltonian of ambient thermal phonons with its formula as follows,

\begin{align}
\hat{H}_{photon}=\sum_{q}\hbar\omega_{q}\hat{a}^{\dag}_{q}\hat{a}_{q},\label{9}
\end{align}

\noindent where $\hat{a}^{\dag}_{q}$,$\hat{a}_{q}$ are the photon creation, annihilation operators with its q-th noninteracting photon mode frequency $\omega_{q}$. Considering the weak coupling, the last item $\hat{H}_{V}$ in Eq.(\ref{4}) represents the interaction between the system and the ambient thermal phonons, with the corresponding coupling strength as follows,

\begin{eqnarray}
&\hat{H}_{V}=&\sum_{i=1,2}\sum_{q}(\varepsilon_{ci}\hat{\sigma}_{ci}\otimes\hat b_{\gamma_{ci}q}^{\dag}+\varepsilon_{ci}^{\ast}\hat{\sigma}_{ci}^{\dag}\otimes\hat b_{\gamma_{ci}q})\nonumber\\
            && +\sum_{q} (\varepsilon_{0v}\hat{\sigma}_{0v}\otimes\hat b_{\Gamma_{R}q}^{\dag}+\varepsilon_{0v}^{\ast}\hat{\sigma}_{0v}^{\dag}\otimes\hat b_{\Gamma_{R}q}).    \label{10}
\end{eqnarray}

\noindent In Eq.(\ref{10}), $\sigma_{ci}$=$|c\rangle\langle i|_{(i=1,2)}$, $\sigma_{0v}$=$|0\rangle\langle v|$ are the Pauli fall operators corresponding to their transition processes.   $\hat{b}^{\dag}_{ij}$,$\hat{b}_{ij}$ are the phonon creation, annihilation operators, and $\varepsilon_{ij}$ is the coupling coefficient.

Under the Born-Markov and Weisskopf-Wigner approximations\cite{WANG1974323,zhao2020_high,zhong_2021_photovoltaic} in the Schr$\ddot{o}$dinger picture, the dynamics of the reduced density operators of the system can be written as,

\begin{align}
\frac{d\hat{\rho}}{dt}=-i[\hat{H}_{S},\hat{\rho}]+\mathscr{L}\hat{\rho},\label{11}
\end{align}

\noindent where $\hat{\rho}$ is the reduced density operators, and  the dissipative $\mathscr{L}\hat{\rho}$ is decomposed into the following components,

\begin{equation}
\mathscr{L}\hat{\rho}=\sum_{i=1,2}\mathscr{L}_{\gamma_{0i}}\hat{\rho}+\sum_{i=1,2}\mathscr{L}_{ic}\hat{\rho}+\mathscr{L}_{\Gamma_{R}}\hat{\rho}+\mathscr{L}_{\Gamma}\hat{\rho},\label{12}
\end{equation}

\noindent The components in Eq.(\ref{12}) are expressed as follows,

\begin{eqnarray}
&\mathscr{L}_{\gamma_{0i}}\hat{\rho}(i\!=\!1,2)&\!=\!\frac{\gamma_{0i}}{2}[(n_{0i}+1)(2\hat{\sigma}_{0i}\hat{\rho}\hat{\sigma}_{0i}^{\dag}-\hat{\sigma}_{0i}^{\dag}\hat{\sigma}_{0i}\hat{\rho}-\hat{\rho}\hat{\sigma}_{0i}^{\dag}\hat{\sigma}_{0i})\nonumber\\
                                                &&+n_{0i}(2\hat{\sigma}_{0i}^{\dag}\hat{\rho}\hat{\sigma}_{0i}-\hat{\sigma}_{0i}\hat{\sigma}_{0i}^{\dag}\hat{\rho}-\hat{\rho}\hat{\sigma}_{0i}\hat{\sigma}_{0i}^{\dag})],\label{13}\\
&\mathscr{L}_{\gamma_{ci}}\hat{\rho}(i\!=\!1,2)&\!=\!\frac{\gamma_{ci}}{2}[(n_{ic}+1)(2\hat{\sigma}_{ci}\hat{\rho}\hat{\sigma}_{ci}^{\dag}-\hat{\sigma}_{ci}^{\dag}\hat{\sigma}_{ci}\hat{\rho}-\hat{\rho}\hat{\sigma}_{ci}^{\dag}\hat{\sigma}_{ci})\nonumber\\
                                                &&+n_{ci}(2\hat{\sigma}_{ci}^{\dag}\hat{\rho}\hat{\sigma}_{ci}-\hat{\sigma}_{ci}\hat{\sigma}_{ci}^{\dag}\hat{\rho}-\hat{\rho}\hat{\sigma}_{ci}\hat{\sigma}_{ci}^{\dag})],\label{14}\\
&\mathscr{L}_{\Gamma_{R}}\hat{\rho}&\!=\!\frac{\Gamma_{R}}{2}[(N_{c}+1)(2\hat{\sigma}_{0v}\hat{\rho}\hat{\sigma}_{0v}-\hat{\sigma}_{0v}^{\dag}\hat{\sigma}_{0v}\hat{\rho}-\hat{\rho}\hat{\sigma}_{0v}^{\dag}\hat{\sigma}_{0v})\nonumber\\
                                                &&+N_{c}(2\hat{\sigma}_{0v}^{\dag}\hat{\rho}\hat{\sigma}_{0v}-\hat{\sigma}_{0v}\hat{\sigma}_{0v}^{\dag}\hat{\rho}-\hat{\rho}\hat{\sigma}_{0v}\hat{\sigma}_{0v}^{\dag})],\label{15}\\
&\mathscr{L}_{\Gamma}\hat{\rho}&\!=\!\frac{\Gamma}{2}(2\hat{\sigma}_{vc}\hat{\rho}\hat{\sigma}_{vc}^{\dag}-\hat{\sigma}_{vc}^{\dag}\hat{\sigma}_{vc}\hat{\rho}-\hat{\rho}\hat{\sigma}_{vc}^{\dag}\hat{\sigma}_{vc})\label{16}
\end{eqnarray}

\noindent As shown in Fig.\ref{Fig.1}, the dissipative process in the transition $|0\rangle$ $\leftrightarrow$ $|i\rangle$ driven by photon pulses is described by $\mathscr{L}_{\gamma_{0i}}\hat{\rho}$ with spontaneous decay rates $\gamma_{0i}$, Pauli operator $\hat{\sigma}_{0i}$=$|0\rangle\langle i|$, $\hat{\sigma}_{0i}^{\dag}$=$|i\rangle\langle 0|$ and $n_{0i}$ being the average occupation numbers. At ambient temperature $T_{a}$, the dissipative  transition $|i\rangle$ $\leftrightarrow$ $|c\rangle$ is represented by $\mathscr{L}_{\gamma_{ci}}$ with spontaneous decay rates $\gamma_{ci}$, where $\hat{\sigma}_{ci}$=$|c\rangle\langle i|$, $\hat{\sigma}_{ci}^{\dag}$=$|i\rangle\langle c|$, and the corresponding phonon occupation numbers  \(n_{ic}$=$[exp(\frac{(E_{i}-E_{c})}{k_{B}T_{a}})-1]^{-1}(i$=$1,2)\). Similarly, $\mathscr{L}_{\gamma_{0v}}$ denotes another interaction between the system and the ambient thermal phonons via the transition $|v\rangle$ $\leftrightarrow$ $|0\rangle$ with $\hat{\sigma}_{0v}$=$|0\rangle\langle v|$, $\hat{\sigma}_{0v}^{\dag}$=$|v\rangle\langle 0|$, where \(N_{c}$=$[exp(\frac{(E_{v}-E_{0})}{k_{B}T_{a}})-1]^{-1}\) denotes the corresponding average phonon occupation numbers, and the transition rate is $\Gamma_{R}$. The last term $\mathscr{L}_{\Gamma}$ describes the transition $|c\rangle$ $\rightarrow$ $|v\rangle$ with the relaxation rate $\Gamma$, and the operator $\hat{\sigma}_{vc}$ is defined as $\hat{\sigma}_{vc}=|v\rangle\langle c|$. Here, $|c\rangle$ and $|v\rangle$ can be treated as charge separation states, and the output electric current is proportional to the relaxation rate $\Gamma$, $j$=$e\Gamma\rho_{cc}$.

Therefore, the dynamics of the reduced density matrix elements $\hat{\rho}$ can be written in according with Eq.(\ref{11})$\sim$ Eq.(\ref{16}) as follows,

\begin{widetext}
\begin{eqnarray}
&\dot{\rho}_{11}\!=\!&-i\Omega(\rho_{21}-\rho_{12})-\sqrt{\gamma_{01}}g_{1}(t)\rho_{01}-\sqrt{\gamma_{01}}g^{\ast}_{1}(t)\rho_{10}-\gamma_{01}[(n_{01}+1)\rho_{11}-n_{01}\rho_{00}] \nonumber\\
                    &&-\gamma_{1c}[(n_{1c}+1)\rho_{11}-n_{1c}\rho_{cc}],\label{17}\\
&\dot{\rho}_{22}\!=\!&-i\Omega(\rho_{12}-\rho_{21})-\sqrt{\gamma_{02}}g_{2}(t)\rho_{02}-\sqrt{\gamma_{02}}g^{\ast}_{2}(t)\rho_{20}-\gamma_{02}[(n_{02}+1)\rho_{22}-n_{02}\rho_{00}] \nonumber\\
                        &&-\gamma_{2c}[(n_{2c}+1)\rho_{22}-n_{2c}\rho_{cc}],\label{18}\\
&\dot{\rho}_{12}\!=\!&-i\varepsilon\rho_{12}-i\Omega(\rho_{22}-\rho_{11})-\sqrt{\gamma_{01}}g_{1}(t)\rho_{02}-\sqrt{\gamma_{02}}g^{\ast}_{2}(t)\rho_{10}-\frac{\rho_{12}}{2}[\gamma_{01}(n_{01}+1) \nonumber\\
                       &&+\gamma_{02}(n_{02}+1)+\gamma_{1c}(n_{1c}+1)+\gamma_{2c}(n_{2c}+1)],\\                                                                \label{19}
&\dot{\rho}_{21}\!=\!&i\varepsilon\rho_{21}+i\Omega(\rho_{22}-\rho_{11})-\sqrt{\gamma_{02}}g_{2}(t)\rho_{01}-\sqrt{\gamma_{01}}g^{\ast}_{1}(t)\rho_{20}-\frac{\rho_{21}}{2}[\gamma_{01}(n_{01}+1) \nonumber\\
                      &&+\gamma_{02}(n_{02}+1)+\gamma_{1c}(n_{1c}+1)+\gamma_{2c}(n_{2c}+1)],\\                                                                  \label{20}
&\dot{\rho}_{01}\!=\!&i\frac{\varepsilon}{2}\rho_{01}+i\Omega\rho_{02}+\sqrt{\gamma_{01}}g^{\ast}_{1}(t)(\rho_{11}-\rho_{00})+\sqrt{\gamma_{02}}g^{\ast}_{2}(t)\rho_{21}-\frac{\rho_{01}}{2}[\gamma_{01}(n_{01}+1) \nonumber\\
                      &&+\gamma_{01}n_{01}+\gamma_{02}n_{02}+\gamma_{1c}(n_{1c}+1)+\Gamma_{R}N_{c}],\\                                                            \label{21}
&\dot{\rho}_{10}\!=\!&-i\frac{\varepsilon}{2}\rho_{10}-i\Omega\rho_{20}+\sqrt{\gamma_{01}}g_{1}(t)(\rho_{11}-\rho_{00})+\sqrt{\gamma_{02}}g_{2}(t)\rho_{12}-\frac{\rho_{10}}{2}[\gamma_{01}(n_{01}+1)\nonumber\\
                       &&+\gamma_{01}n_{01}+\gamma_{02}n_{02}+\gamma_{1c}(n_{1c}+1)+\Gamma_{R}N_{c}],\\                                                           \label{22}
&\dot{\rho}_{02}\!=\!&-i\frac{\varepsilon}{2}\rho_{02}+i\Omega\rho_{01}+\sqrt{\gamma_{02}}g^{\ast}_{2}(t)(\rho_{22}-\rho_{00})+\sqrt{\gamma_{01}}g^{\ast}_{1}(t)\rho_{12}-\frac{\rho_{02}}{2}[\gamma_{01}n_{01}+\gamma_{02}(n_{02}+1)\nonumber\\
                       &&+\gamma_{02}n_{02}+\gamma_{2c}(n_{2c}+1)+\Gamma_{R}N_{c}],\\                                                                              \label{23}
&\dot{\rho}_{20}\!=\!&i\frac{\varepsilon}{2}\rho_{20}-i\Omega\rho_{10}+\sqrt{\gamma_{02}}g_{2}(t)(\rho_{22}-\rho_{00})+\sqrt{\gamma_{01}}g_{1}(t)\rho_{21}-\frac{\rho_{20}}{2}[\gamma_{01}n_{01}+\gamma_{02}(n_{02}+1) \nonumber\\
                      &&+\gamma_{02}n_{02}+\gamma_{2c}(n_{2c}+1)+\Gamma_{R}N_{c}],\\                                                                                \label{24}
&\dot{\rho}_{cc}\!=\!&-\Gamma\rho_{cc}+\gamma_{1c}[(n_{1c}+1)\rho_{11}-n_{1c}\rho_{cc}]+\gamma_{2c}[(n_{2c}+1)\rho_{22}-n_{2c}\rho_{cc}],\\     \label{25}
&\dot{\rho}_{vv}\!=\!&-\Gamma_{R}(N_{c}+1)\rho_{vv}+\Gamma_{R}N_{c}\rho_{00}+\Gamma\rho_{cc}.                                                                     \label{26}
\end{eqnarray}
\end{widetext}

\noindent where $\rho_{ii}$ describe the diagonal elements and $\rho_{ij}$ is the non-diagonal elements of the corresponding states. Thus, with the help of the deduced reduced density elements, the output voltage defined as the chemical potential difference between $|c\rangle$ and $|v\rangle$, can be expressed as $eV$=$E_{c}-E_{v}+K_{B}T_{a}\ln\frac{\rho_{cc}}{\rho_{vv}}$, with \(e\) being the fundamental charge. Based on the electric current and voltage, we can easily obtain the power output $P$=$jV$. And the photoelectric conversion efficiency can be measured by $\eta=\frac{P}{P_{in}}$ in this DQD photovoltaic system with

\begin{widetext}
\begin{align}
 P_{in}(t)={}&Tr_{D}[{\rho_{D}\dot{H}_{D}(t)}]=-\dot{g}_{1}(t)\rho_{00}-\dot{g}_{2}(t)\rho_{00}+\dot{g}^{\ast}_{1}(t)\rho_{11}i\dot{g}^{\ast}_{2}(t)\rho_{22}, \label{27}
\end{align}
\end{widetext}

\noindent The equation (\ref{27}) indicates that the output power of the donor acts as the incident power owing to the absorbed incident photons by the donor in our proposed DQD photocell model. We now examine how do the quantum thermodynamic quantities evolve. Accordingly, the quantum thermodynamic quantities of this proposed DQD photocell can be deduced by the reduced density elements according to Eq.(\ref{1}) and Eq.(\ref{3}). Thus, the heat currents of the donor and acceptor in this DQD photocell $\dot{Q}_{D}(t)$ and $\dot{Q}_{A}(t)$ together the entropy flow $\dot{S}_{A}$ of the acceptor can be derived as follows, respectively.

\begin{widetext}
\begin{eqnarray}
&\dot{Q}_{D}(t)\!=\!&Tr_{D}[\dot{\rho}_{D}(t)H_{D}(t)]=\rho_{11}\varepsilon[-\frac{{\gamma_{1c}}}{2}(n_{1c}+1)-\frac{{\gamma_{01}}}{2}(n_{01}+1)] +\rho_{22}\varepsilon[\frac{{\gamma_{2c}}}{2}(n_{2c}+1)+\frac{{\gamma_{02}}}{2}(n_{02}+1)]\nonumber\\{}
&&+\rho_{12}\Omega[-\frac{{\gamma_{1c}}}{2}(n_{1c}+1)-\frac{{\gamma_{2c}}}{2}(n_{2c}+1)-\frac{{\gamma_{01}}}{2}(n_{01}+1)-\frac{{\gamma_{02}}}{2}(n_{02}+1)]\nonumber\\{}
&&+ \rho_{21}\Omega[-\frac{{\gamma_{1c}}}{2}(n_{1c}+1)-\frac{{\gamma_{2c}}}{2}(n_{2c}+1)-\frac{{\gamma_{01}}}{2}(n_{01}+1)-\frac{{\gamma_{02}}}{2}(n_{02}+1)]+\rho_{00}\varepsilon[\frac{{\gamma_{01}}}{2}n_{01}\nonumber\\{}
&&-\frac{{\gamma_{02}}}{2}n_{02}]+\rho_{cc}\varepsilon[\frac{{\gamma_{1c}}}{2}n_{1c}-\frac{{\gamma_{2c}}}{2}n_{2c}] +\rho_{10}[-i\frac{\Gamma_{R}}{2}N_{c}\sqrt{\gamma_{01}}g^{\ast}_{1}(t)\nonumber\\{}
&&-i\frac{\gamma_{1c}}{2}(n_{1c}+1)\sqrt{\gamma_{01}}g^{\ast}_{1}(t)-i\frac{\gamma_{01}}{2}(n_{01}+1)\sqrt{\gamma_{01}}g^{\ast}_{1}(t)-i\frac{\gamma_{01}}{2}n_{01}\sqrt{\gamma_{01}}g^{\ast}_{1}(t) \label{28}\\
&& -i\frac{\gamma_{02}}{2}n_{02}\sqrt{\gamma_{01}}g^{\ast}_{1}(t)]+\rho_{01}[i\frac{\Gamma_{R}}{2}N_{c}\sqrt{\gamma_{01}}g_{1}(t)+i\frac{\gamma_{1c}}{2}(n_{1c}+1)\sqrt{\gamma_{01}}g_{1}(t) \nonumber \\{}
&&+i\frac{\gamma_{01}}{2}(n_{01}+1)\sqrt{\gamma_{01}}g_{1}(t)+i\frac{\gamma_{01}}{2}n_{01}\sqrt{\gamma_{01}}g_{1}(t)+i\frac{\gamma_{02}}{2}n_{02}\sqrt{\gamma_{01}}g_{1}(t)]+\rho_{20}[-i\frac{\Gamma_{R}}{2}N_{c}\sqrt{\gamma_{02}}g^{\ast}_{2}(t) \nonumber{}\\
&&-i\frac{\gamma_{2c}}{2}(n_{2c}+1)\sqrt{\gamma_{02}}g^{\ast}_{2}(t)-i\frac{\gamma_{01}}{2}n_{01}\sqrt{\gamma_{02}}g^{\ast}_{2}(t)-i\frac{\gamma_{02}}{2}n_{02}\sqrt{\gamma_{02}}g^{\ast}_{2}(t)-i\frac{\gamma_{02}}{2}(n_{02}+1)\sqrt{\gamma_{02}}g^{\ast}_{2}(t)]\nonumber\\
&&+\rho_{02}[i\frac{\Gamma_{R}}{2}N_{c}\sqrt{\gamma_{02}}g_{2}(t)+i\frac{\gamma_{2c}}{2}(n_{2c}+1)\sqrt{\gamma_{02}}g_{2}(t)+i\frac{\gamma_{01}}{2}n_{01}\sqrt{\gamma_{02}}g_{2}(t)+i\frac{\gamma_{02}}{2}n_{02}\sqrt{\gamma_{02}}g_{2}(t)\nonumber\\
&&+i\frac{\gamma_{02}}{2}(n_{02}+1)\sqrt{\gamma_{02}}g_{2}(t)],\nonumber{}\\
&\dot{Q}_{A}(t)\!=\!&Tr_{A}[\dot{\rho}_{A}(t)H_{A}(t)]=\frac{{\Gamma_{R}}{\omega_{0}}}{2}(N_{c}+1)\rho_{vv}-\frac{{\Gamma_{R}}{\omega_{0}}}{2}N_{c}\rho_{00}-\Gamma\omega_{0}\rho_{cc}+\frac{{\omega_{0}}{\gamma_{1c}}}{2}(n_{1c}+1)\rho_{11}+\frac{{\omega_{0}}{\gamma_{2c}}}{2}(n_{2c}+1)\rho_{22}\nonumber\\{}
&&-\frac{\omega_{0}}{2}(\gamma_{1c}n_{1c}+\gamma_{2c}n_{2c})\rho_{cc},                            \label{29}\\
&\dot{S}_{A}(t\!=\!&-Tr_{A}[\dot{\rho}(ln\rho)]=(-\Gamma\rho_{cc}+\gamma_{1c}[(n_{1c}+1)\rho_{11}-\gamma_{1c}n_{1c}\rho_{cc}+\gamma_{2c}[(n_{2c}+1)\rho_{22}-\gamma_{2c}n_{2c}\rho_{cc}])ln\rho_{cc}\nonumber\\{}
&&+(1-\Gamma_{R}(N_{c})\rho_{vv}+\Gamma_{R}N_{c}\rho_{00}+\Gamma\rho_{cc})ln\rho_{vv}).            \label{30}
\end{eqnarray}
\end{widetext}

\section{Results and discussion}

\begin{figure}[htp]
\center
\includegraphics[scale=.55]{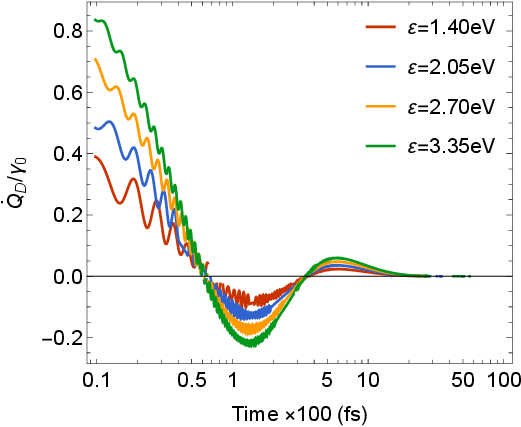}\hspace{0in}%
\includegraphics[scale=.55]{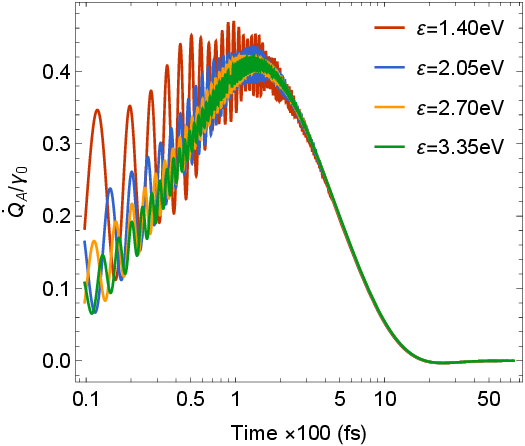}\hspace{0in}%
\includegraphics[scale=.55]{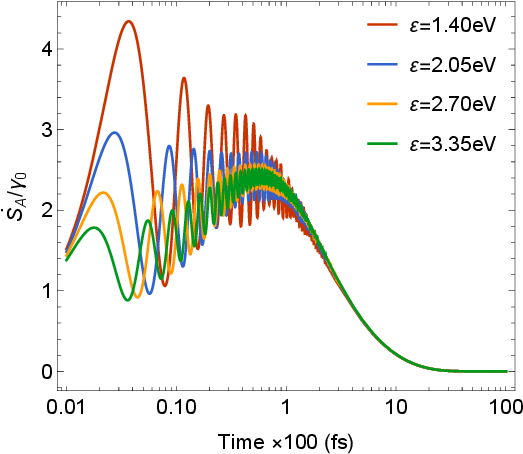}
\caption{(Color online) Dynamics evolution of the heat currents from the environment into the donor \(\dot{Q}_{D}\) and from the donor into accepter \(\dot{Q}_{A}\), and the entropy flow \(\dot{S}_{A}\) of the acceptor in the DQD photocell system influenced by its different gap energies \(\varepsilon\) with other parameters taken from Table \ref{Table 1}.}
\label{Fig.2}
\end{figure}

\begin{figure}[htp]
\center
\includegraphics[scale=.65]{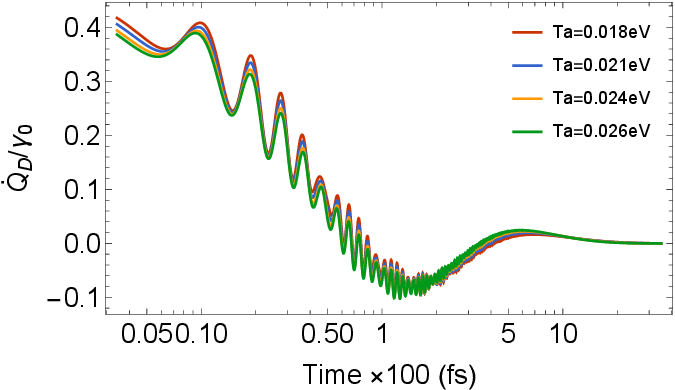 }\hspace{0in}%
\includegraphics[scale=.65]{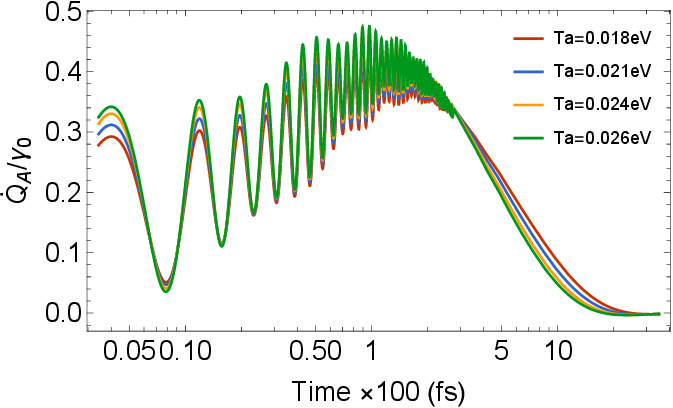 }\hspace{0in}%
\includegraphics[scale=.65]{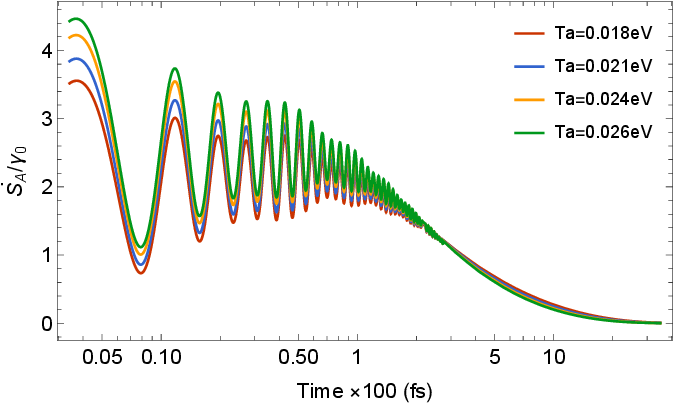 }
\caption{(Color online) Dynamics evolution of the heat currents from the environment into the donor \(\dot{Q}_{D}\) and from the donor into accepter \(\dot{Q}_{A}\), and the entropy flow \(\dot{S}_{A}\) of the acceptor in the DQD photocell system influenced by different ambient temperatures \(T_{a}\) with other parameters taken from Table \ref{Table 1}.}
\label{Fig.3}
\end{figure}

This work will explore the evolution characteristics of the quantum thermodynamic quantities during the photoelectric conversion in the DQD photocells, and evaluates the photovoltaic features under the same physical environment. The goal is attempts to formulate a novel theoretical approach to evaluate DQD photocells performances. Therefore, the heat currents from the donor and acceptor caught our attention. For the direction of heat currents, we use a uniform standard that specifies positive inflow and negative outflow. For the sake of convenience, several typical parameters should be selected before the analysis. The gaps of the two QDS arranged at left and right sides are set to the same value. Considering the photon pulse weakly coupling to the QD photocell, the absorbed photon number in unit time by the left and right QD is $n_{01}$=5, $n_{02}$=6, respectively. And other parameters used here are listed in Table \ref{Table 1}. Fig.\ref{Fig.2} and Fig.\ref{Fig.3} describe the time-evolution of the thermodynamic-related quantities, i.e., the heat currents of the donor \(\dot{Q}_{D}\), the acceptor \(\dot{Q}_{A}\) and the entropy flow \(\dot{S}_{A}\) of the acceptor in the DQD photocell system, for the goal of evaluating the thermodynamic behavior of the photovoltaic process.

\begin{table}
\begin{center}
\caption{Parameters used in the numerical calculations.}
\label{Table 1}
\vskip 0.2cm\setlength{\tabcolsep}{0.5cm}
\begin{tabular}{ccc}
\hline
\hline
                                                                    & Values                     &Units\\
\hline
\(\gamma_{01}\)                                                     & 1.1 \(\gamma_{0}\)         & eV  \\
\(\gamma_{02}\)                                                     & 0.8 \(\gamma_{0}\)         & eV  \\
\(\gamma_{1c}\)                                                     & 1.05 \(\gamma_{0}\)        & eV  \\
\(\gamma_{2c}\)                                                     & 1.1 \(\gamma_{0}\)         & eV  \\
\(\Gamma\)                                                          & 0.16\(\gamma_{0}\)         & eV  \\
\(\Gamma_{R}\)                                                      & 1.25 \(\gamma_{0}\)        & eV  \\
\(\omega_{0}\)                                                      & 1.2                        & eV  \\
$\Omega$                                                            & 0.09                       &     \\
$E_{1}-E_{c}$                                                       & 0.02                       & eV  \\
$E_{2}-E_{c}$                                                       & 0.05                       & eV  \\
$E_{v}-E_{0}$                                                       & 0.08                       & eV  \\
$g^{\ast}_{1}(t)$=$g_{1}(t)$                                        & 0.2                        &     \\
$g^{\ast}_{2}(t)$=$g_{2}(t)$                                        & 0.105                       &     \\
\(\gamma_{0}\)                                                      & \(10^{-3}\)                & scale unit \\
\hline
\hline
\end{tabular}
\end{center}
\end{table}

Fig.\ref{Fig.2} shows the dynamics evolutions of the heat currents \(\dot{Q}_{D}\), \(\dot{Q}_{A}\) and the entropy flow \(\dot{S}_{A}\) of the acceptor with different gap energies $\varepsilon$ when the DQD photocell system operates at an ambient temperature $T_{a}=0.025eV$. As the curves shown in Fig.\ref{Fig.2}, the heat currents \(\dot{Q}_{D}\) and \(\dot{Q}_{A}\) show the in-phase oscillation as the gap energy \(\varepsilon\) increases by 0.65$eV$. The curves of \(\dot{Q}_{D}\) illustrate that the excited donor becomes decayed by the photon pulse during the time interval of [0, 50$fs$], which indicates the gradually decreasing input energy from the photon pulse. In the time interval [50, 350$fs$], the peak heat current \(\dot{Q}_{D}\) appears which indicates that more energy will be absorbed by the acceptor. The curves of \(\dot{Q}_{A}\) demonstrated this conclusion in the same time interval. Not only that, we noticed that the peaks in the time interval of [0, 50$fs$] increase with the gap energies $\varepsilon$, for instance, the curve with $\varepsilon$=3.35$eV$. This means more photons will be absorbed, which is coincided with the conclusion from Ref.\cite{Zhao_2019}. As time goes on, when t is greater than 3000$fs$, both \(\dot{Q}_{D}\) and \(\dot{Q}_{A}\) become decayed to zero, the donor falls in thermal equilibrium with the acceptor. And the dynamic entropy flow \(\dot{S}_{A}\) of the acceptor in Fig.\ref{Fig.2} proves this physical phenomena.

The effect of fluctuation in ambient temperature near room temperature ($T_{a}$$=$0.0259$eV$) on thermodynamic quantities merits additional investigation. Hence, the curves for the heat currents \(\dot{Q}_{D}\), \(\dot{Q}_{A}\) and the entropy flow \(\dot{S}_{A}\) of the acceptor are shown in Fig.\ref{Fig.3} with different ambient temperatures $T_{a}$ with the energy gap $\varepsilon$=2.05$eV$ and other parameters selected from Table \ref{Table 1}. The curves for heat currents \(\dot{Q}_{D}\) show a decayed bipartite oscillation over the time, and then they asymptotically goes to zero, which indicates that the donor in the QDQ photocell becomes in thermal equilibrium with the input photon pulse in the long time limit.
The oscillation amplitudes of \(\dot{Q}_{D}\) illustrate that $T_{a}$ thread a negative influence on the heat currents \(\dot{Q}_{D}\) from the donor, the read curve demonstrates this with the least $T_{a}$$=$0.018$eV$.
However, the evolution of \(\dot{Q}_{A}\) presents two unique aspects in Fig.\ref{Fig.3}. One is the oscillation amplitudes of \(\dot{Q}_{A}\) increases with the $T_{a}$, i.e., a positive role of $T_{a}$ in the evolution of \(\dot{Q}_{A}\). And we see the least oscillation amplitude with $T_{a}$$=$0.018$eV$, the largest oscillation amplitude with $T_{a}$$=$0.026$eV$ from the curves of \(\dot{Q}_{A}\).  The other is the the acceptor becomes in equilibrium with donor is later than that of \(\dot{Q}_{D}\), which can be drawn from the asymptotic zero in the curves of \(\dot{Q}_{A}\) with later time. This conclusion is confirmed by the dynamic entropy flow \(\dot{S}_{A}\) of receptors in Fig.\ref{Fig.3} during the same time domain.

\begin{figure}[htp]
\center
\includegraphics[scale=.55]{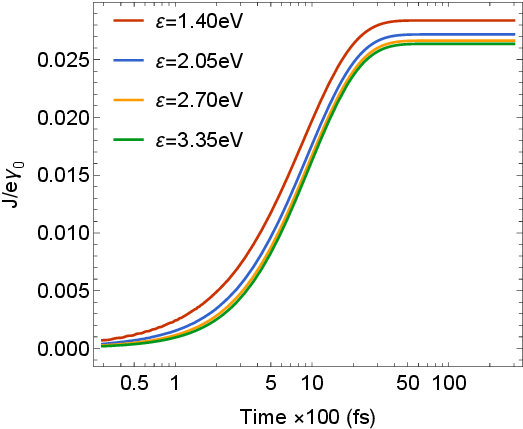 }\hspace{0in}%
\includegraphics[scale=.55]{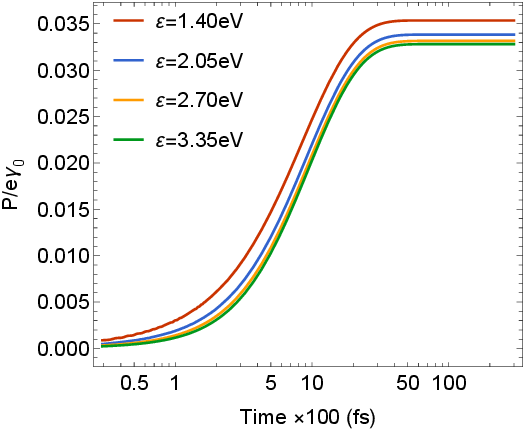 }\hspace{0in}%
\includegraphics[scale=.55]{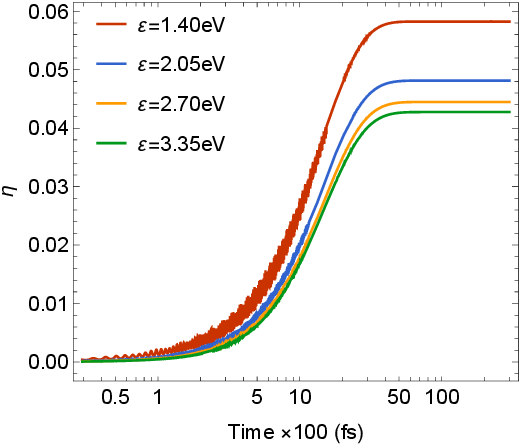 }
\caption{(Color online) Dynamics evolution of photovoltaic current \(J\), power \(P\) and photoelectric conversion efficiency \(\eta\) with different gap energies \(\varepsilon\). Other parameters are the same to those in Fig.\ref{Fig.2}.}
\label{Fig.4}
\end{figure}

\begin{figure}[htp]
\center
\includegraphics[scale=.65]{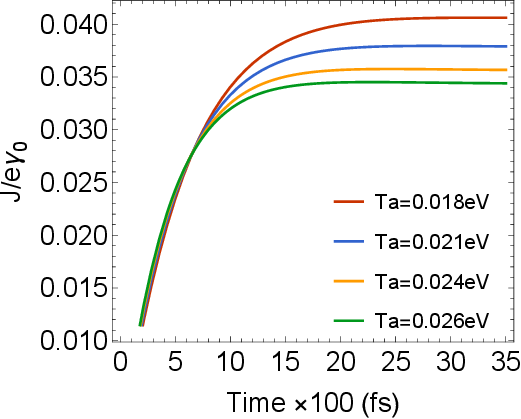 }\hspace{0in}%
\includegraphics[scale=.65]{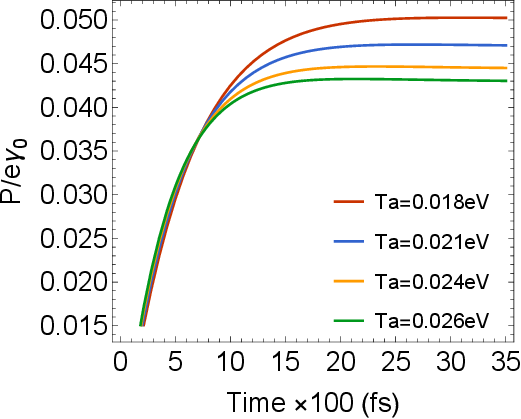 }\hspace{0in}%
\includegraphics[scale=.60]{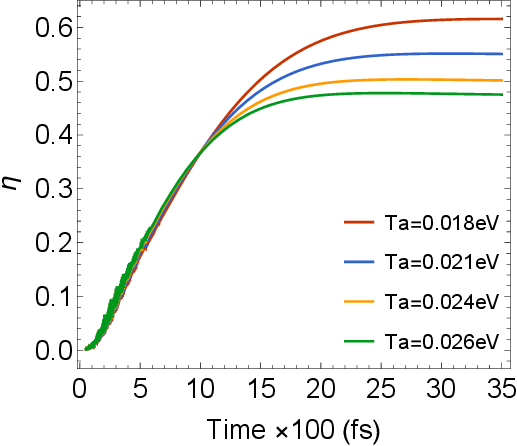 }
\caption{(Color online)Dynamics evolution of photovoltaic current \(J\), power \(P\) and photoelectric conversion efficiency \(\eta\)  with ambient temperatures \(T_{a}\).  Other parameters are the same to those in Fig.\ref{Fig.3}.}
\label{Fig.5}
\end{figure}

We now examine how the photovoltaic characteristics depend on the gap energies \(\varepsilon\) and ambient temperatures \(T_{a}\) in Fig.\ref{Fig.4} and  Fig.\ref{Fig.5}, respectively. The electric current evolves in the same parameter situations with Fig.\ref{Fig.2}, except $n_{01}$=0.5, $n_{02}$=0.6, respectively. From the curves in Fig.\ref{Fig.4}, we see that the gap energies \(\varepsilon\) play a negative role in the photovoltaic current, i.e., the electric current $j$ decreases with the increment of \(\varepsilon\). But the horizontal lines indicate the moment of steady output current is not affected by the increment of \(\varepsilon\), it still maintains around 3500 $fs$ when \(\varepsilon\) is different. The output power $P$ and the photoelectric conversion efficiency \(\eta\) behave similarly to the electric current $j$ in Fig.\ref{Fig.4}. By comparing the curves in Fig.\ref{Fig.2} and Fig.\ref{Fig.4}, it is easy to conclude that the moment of thermodynamic equilibrium is earlier than that of steady-state photovoltaic output characteristics.

The photovoltaic characteristics dependent ambient temperatures $T_{a}$ is plotted in Fig.\ref{Fig.5}. As shown by the curves of current $j$ in Fig.\ref{Fig.5}, we observed that the higher the ambient temperature, the quicker the time it takes for the photovoltaic system to steady state. The higher the surrounding temperature, however, the lower the steady output current. And the effect of temperature $T_{a}$ on the evolution of output power $P$ and on the photoelectric conversion efficiency \(\eta\) are in a similar way to the electric current $j$ in Fig.\ref{Fig.5}.

In this work, we examined the thermodynamic evolution and photoelectric transport properties of DQD photocell under identical parameter settings. When we compared the thermodynamic evolution characteristics of the donor and acceptor, as well as the moments when they reach thermodynamic equilibrium in Fig.\ref{Fig.2} and  Fig.\ref{Fig.3}, to the electrical parameters' steady-state properties of photovoltaic system in Fig.\ref{Fig.4} and  Fig.\ref{Fig.5}, we concluded some interesting phenomena: First, the DQD photocell system's thermodynamic equilibrium behavior comes before its electrical behavior. Second, when the gap energy is regulated, the thermodynamic evolution property of the acceptor follows the same pattern as the output's electrical performance in this photovoltaic system; third, when the ambient temperature is regulated, the acceptor's thermodynamic evolution properties follows an opposite law of increment and decrement in this photovoltaic system. This conclusions imply that we may optimize the photovoltaic performance via some thermodynamically related parameters, resulting in a novel technique for achieving efficient photovoltaic performance. For example, by selecting the proper gap energy at the suitable ambient temperature for a DQD photocell, the electrical features can be indirectly reflected. The merit of this work may be that it shows some thermodynamic quantities to evaluate photovoltaic performance in a DQD photocell.

\section{Conclusions}

To summarize, in a donor-acceptor DQD photocell model driven by a photon pulse, we attempt to formulate a novel theoretical approach to evaluate its photovoltaic performance in this work. The results show that the DQD photocell system's thermodynamic equilibrium behavior goes ahead its electrical behavior. When the gap energy is regulated, the thermodynamic evolution properties of the acceptor follows the same pattern of increase and decrease to the output's electrical performance in DQD photocell system. But, the acceptor's thermodynamic evolution properties follow an opposite law when the ambient temperature is regulated in this photovoltaic system. We speculate that the thermodynamic characteristics may be used to evaluate its photovoltaic performance in a donor-acceptor DQD photocell model. Furthermore, we believe it will help to evaluate photovoltaic qualities by examining the thermodynamic evolution of the acceptor.

\section{acknowledgments}

We offer our thanks for the financial support from the National Natural Science Foundation of China (Grant Nos. 62065009 and 61565008), and Yunnan Fundamental Research Projects, China (Grant No. 2016FB009).

\section*{Conflict of Interest}

The authors declare that they have no conflict of interest. This article does not contain any studies with human participants or animals performed by any of the authors. Informed consent was obtained from all individual participants included in the study.

\bibliography{reference}

\begin{thebibliography}{10}

\bibitem{li_2012_icolloquiumi}
N.~Li, J.~Ren, L.~Wang, G.~Zhang, P.~Hnggi, and B.~Li.
\newblock Colloquium: Phononics: Manipulating heat flow with electronic analogs
  and beyond.
\newblock {\em Reviews of Modern Physics}, 84:1045--1066, 07 2012.

\bibitem{li_2021_influence}
L.~F. Li and S.~C. Zhao.
\newblock Influence of the coupled-dipoles on photosynthetic performance in a
  photosynthetic quantum heat engine.
\newblock {\em Chin. Phys. B}, 30:044215, 04 2021.

\bibitem{1-xu2022Photosynthetic}
L.~X. Xu, S.~C. Zhao, and L.~F. Li.
\newblock Photosynthetic properties assisted by the quantum entanglement in two
  adjacent pigment molecules.
\newblock {\em European Phys. J. Plus}, 136:683, 2022.

\bibitem{4-zhongzhao2021}
S.~Q. Zhong, S.~C. Zhao, and S.~N. Zhu.
\newblock Photovoltaic properties enhanced by the tunneling effect in a coupled
  quantum dot photocell.
\newblock {\em Results in Physics}, 24(4):104094, 2021.

\bibitem{gardiner_1991_quantum}
C.~W. Gardiner.
\newblock {\em Quantum noise}.
\newblock Berlin Heidelberg New York London Paris Tokyo Hong Kong Barcelona
  Budapest Springer, 1991.

\bibitem{heinzpeterbreuer_2006_the}
H.~P Breuer and F.~Petruccione.
\newblock {\em The theory of open quantum systems}.
\newblock Clarendon Press ; Oxford, 2006.

\bibitem{scully2011quantum}
M.~O. Scully, K.~R. Chapin, K.~E. Dorfman, M.~B. Kim, and A.~Svidzinsky.
\newblock Quantum heat engine power can be increased by noise-induced
  coherence.
\newblock {\em PNAS}, 108:15097--15100, 08 2011.

\bibitem{svidzinsky_2011_enhancing}
A.~A. Svidzinsky, K.~E. Dorfman, and M.~O. Scully.
\newblock Enhancing photovoltaic power by fano-induced coherence.
\newblock {\em Phys. Rev. A}, 84, 11 2011.

\bibitem{chen_2020_radiative}
J.~Y. Chen and S.~C. Zhao.
\newblock Radiative recombination rate suppressed in a quantum photocell with
  three electron donors.
\newblock {\em European Phys. J. Plus}, 135, 01 2020.

\bibitem{zhang_2015_delocalized}
Y.~Zhang, S.~Oh, F.~H. Alharbi, G.~S. Engel, and S.~Kais.
\newblock Delocalized quantum states enhance photocell efficiency.
\newblock {\em Physical Chemistry Chemical Physics}, 17:5743--5750, 2015.

\bibitem{zhong_2021_photovoltaic}
S.~Q. Zhong, S.~C. Zhao, and S.~N. Zhu.
\newblock Photovoltaic performances in a cavity-coupled double quantum dots
  photocell.
\newblock {\em Results in Physics}, 27:104503, 08 2021.

\bibitem{1979The}
R.~Alicki.
\newblock The quantum open system as a model of the heat engine.
\newblock {\em Journal of Physics A: Mathematical and General}, 12:L103, 1979.

\bibitem{cuetara_2016_quantum}
G.~B. Cuetara, M.~Esposito, and G.~Schaller.
\newblock Quantum thermodynamics with degenerate eigenstate coherences.
\newblock {\em Entropy}, 18:447, 12 2016.

\bibitem{PhysRevLett.122.150603}
K.~Ptaszy\ifmmode~\acute{n}\else \'{n}\fi{}ski and M.~Esposito.
\newblock Thermodynamics of quantum information flows.
\newblock {\em Phys. Rev. Lett.}, 122:150603, 2019.

\bibitem{spohn_1978_irreversible}
H.~Spohn and J.~L. Lebowitz.
\newblock {\em Irreversible Thermodynamics for Quantum Systems Weakly Coupled
  to Thermal Reservoirs}, volume~38.
\newblock John Wiley and Sons, Inc., 03 1978.

\bibitem{seifert_2012_stochastic}
U.~Seifert.
\newblock Stochastic thermodynamics, fluctuation theorems and molecular
  machines.
\newblock {\em Reports on Progress in Physics}, 75:126001, 11 2012.

\bibitem{evans_1993_probability}
D.~J. Evans, E.~G.~D. Cohen, and G.~P. Morriss.
\newblock Probability of second law violations in shearing steady states.
\newblock {\em Phys. Rev. Lett.}, 71:2401--2404, 10 1993.

\bibitem{PhysRevA.74.063823}
E.~Boukobza and D.~J. Tannor.
\newblock Thermodynamics of bipartite systems: Application to light-matter
  interactions.
\newblock {\em Phys. Rev. A}, 74, 12 2006.

\bibitem{dong_2021_thermodynamic}
H.~Dong, A.~Ghosh, M.~O. Scully, and G.~Kurizki.
\newblock Thermodynamic bounds on work extraction from photocells and
  photosynthesis.
\newblock {\em European Phys. J. Spec. Topics}, 230:873--879, 04 2021.

\bibitem{Weimer_2008}
H.~Weimer, M.~J. Henrich, F.~Rempp, H.~Schr?der, and G.~Mahler.
\newblock Local effective dynamics of quantum systems: A generalized approach
  to work and heat.
\newblock {\em Europhys. Lett.}, 83(3):30008, 2008.

\bibitem{quan_2007_quantum}
H.~T. Quan, Y.~X. Liu, C.~P. Sun, and F.~Nori.
\newblock Quantum thermodynamic cycles and quantum heat engines.
\newblock {\em Phys. Rev. E}, 76, 9 2007.

\bibitem{1980Thermodynamic}
P.~T. Landsberg and G.~Tonge.
\newblock Thermodynamic energy conversion efficiencies.
\newblock {\em Journal of Applied Physics}, 51(7):R1--R20, 1980.

\bibitem{Spohn1978Entropy}
H.~Spohn.
\newblock Entropy production for quantum dynamical semigroups.
\newblock {\em Journal of Mathematical Physics}, 19:1227--1230, 05 1978.

\bibitem{kadison_1965_mathematical}
R.~V. Kadison and G.~W. Mackey.
\newblock Mathematical foundations of quantum mechanics.
\newblock {\em The American Mathematical Monthly}, 72:96, 01 1965.

\bibitem{2019The}
A.~C. Jorge, M.~C. H¨¦ctor, and A.~Z¨²iga-Segundo.
\newblock The von neumann entropy for mixed states.
\newblock {\em Entropy}, 21(1), 2019.

\bibitem{li2017new}
J.~Li and H.~Cao.
\newblock A new generalization of von neumann relative entropy.
\newblock {\em International Journal of Theoretical Physics}, 56(11):3405,
  2017.

\bibitem{2017General}
J.~A. Caputo, L.~C. Frenette, N.~Zhao, K.~L. Sowers, T.~D. Krauss, and D.~J.
  Weix.
\newblock General and efficient c-c bond forming photoredox catalysis with
  semiconductor quantum dots.
\newblock {\em Journal of the American Chemical Society}, 139(12):4250, 2017.

\bibitem{doi:10.1021/nl101490z}
M.~C. Beard, A.~G. Midgett, M.~C. Hanna, J.~M. Luther, B.~K. Hughes, and A.~J.
  Nozik.
\newblock Comparing multiple exciton generation in quantum dots to impact
  ionization in bulk semiconductors: Implications for enhancement of solar
  energy conversion.
\newblock {\em Nano Letters}, 10(8):3019, 2010.

\bibitem{doi:10.1021/ar3001958}
M.~C. Beard, J.~M. Luther, O.~E. Semonin, and A.~J. Nozik.
\newblock Third generation photovoltaics based on multiple exciton generation
  in quantum confined semiconductors.
\newblock {\em Accounts of Chemical Research}, 46(6):1252, 2013.

\bibitem{2019PbS}
I.~Candan, M.~Parlak, and C.~Ercelebi.
\newblock Pbs quantum dot enhanced p-cigs/n-si heterojunction diode.
\newblock {\em Journal of materials science}, 30(3):2127, 2019.

\bibitem{Fan029}
X.~B. Fan, S.~Yu, B.~Hou, and J.~M. Kim.
\newblock Quantum dots based photocatalytic hydrogen evolution.
\newblock {\em Israel Journal of Chemistry}, 59(8):762, 2019.

\bibitem{Zhao_2019}
S.~C. Zhao and J.~Y. Chen.
\newblock Enhanced quantum yields and efficiency in a quantum dot photocell
  modeled by a multi-level system.
\newblock {\em New Journal of Physics}, 21:103015, 10 2019.

\bibitem{lira_2022_enhanced}
J.~Lira, J.~M. Villas-Boas, L.~Sanz, and A.~M. Alcalde.
\newblock Enhanced solar photocurrent using a quantum-dot molecule.
\newblock {\em J. Opt. Soc. of Am. B}, 39:2047, 07 2022.

\bibitem{loudon2010the}
R.~Loudon.
\newblock {\em The quantum theory of light}.
\newblock Oxford University Press, 2010.

\bibitem{wang2011efficient}
Y.~M. Wang, J.~Min$\acute{¨¢}\breve{r}$, L.~Sheridan, and V.~Scarani.
\newblock Efficient excitation of a two-level atom by a single photon in a
  propagating mode.
\newblock {\em Phys. Rev. A}, 83, 06 2011.

\bibitem{WANG1974323}
Y.~K. Wang and I.~C. Khoo.
\newblock On the wigner-weisskopf approximation in quantum optics.
\newblock {\em Opt. Commun.}, 11(4):323--326, 1974.

\bibitem{zhao2020_high}
S.~C. Zhao and Q.~X. Wu.
\newblock High quantum yields generated by a multi-band quantum dot photocell.
\newblock {\em Superlattices and Microstructures}, 137:106329, 01 2020.

\end{thebibliography}
\bibliographystyle{unsrt}
\end{document}